\documentclass[11pt,thmsa,a4paper,english]{article}
\usepackage{babel}
\RequirePackage{amsthm} \RequirePackage[cmex10]{amsmath}
\RequirePackage{natbib}
\RequirePackage[colorlinks,citecolor=blue,urlcolor=blue]{hyperref}
\usepackage{amssymb,amstext}
\usepackage{booktabs,graphics,fancybox,ulem,color,pifont,amsfonts,amsbsy}
\usepackage{subfigure,epsfig,calc}
\usepackage{eurosym,picture,tikz,amsthm,bm}
\usepackage[dvipdfm]{geometry}
\usepackage{verbatim}
\usepackage{rotating}

\begin{document}

%\begin{frontmatter}
\title{The Africa-Dummy: Gone with the Millennium?}
%\runtitle{No tiltle}
%\thank{Thankx}

%\begin{aug}
\author{Max K\"ohler\footnote{University of G\"ottingen, Courant Research Center for Poverty, 
Equity and Growth, G\"ottingen, Germany},
Stefan Sperlich\footnote{Corresponding author: Universit\'e de Gen\'eve,  Switzerland, stefan.sperlich@unige.ch}}
%\end{aug}

\date{DP version 2014/15}

\maketitle

\begin{abstract} \noindent
A fixed effects regression estimator is introduced that can directly identify and estimate the Africa-Dummy in one regression step so that its correct standard errors as well as correlations to other coefficients can easily be estimated. 
We can estimate the Nickel bias and found it to be negligibly tiny.
Semiparametric extensions check whether the Africa-Dummy is simply a result of misspecification of the functional form. In particular, we show that the returns to growth factors are different for Sub-Saharan 
African countries compared to the rest of the world. For example, returns to population growth are positive and beta-convergence is faster. 
When extending the model to identify the development of the Africa-Dummy over time we see
that it has been changing dramatically over time and that the punishment for Sub-Saharan African countries has been 
decreasing incrementally to reach insignificance around the turn of the millennium.\footnote{We 
acknowledge helpful discussion and comments from Inmaculada Martínez-Zarzoso, Jacques Silber,
Stephan Klasen, and Thomas Kneib as well as some help with data preparation by Julian Vortmeyer.}
\end{abstract}


\newpage

\section{Introduction} \label{sec1}

To study the Africa-Dummy we start out from the classical growth model of \citet{mankiw_1992}. This model contains several simplifications; most countries possess certain characteristics that are hard to measure and to incorporate but represent systematic drivers for growth like for example international capital markets (\citet{barro_1995}).  \citet{islam_1995} criticized that countries have fundamentally differing production functions so that comparisons between their economies are difficult. A further simplification is the assumption that the endowment with resources can be infinitely substituted by capital. \citet{georgescu-roegen_1975} argue that this point of view is too optimistic with respect to the limitations of technological progress. Other variables that are correlated to economic growth but not incorporated in the growth model are political factors (see \citet{collier_1999}), diseases like AIDS (see \citet{were_2003}), geographical factors and trade openness (see \citet{sachs_1997}), ethnic diversity (see \citet{easterly_1997}) or historical background such as the colonial heritage (see \citet{price_2003}), to mention a few. Among others, these problems result in empirical weaknesses. 
Among others, \citet{barossi_2005} summarize that among most regressions the estimated capital share exceeds the value obtained from the national accounts and that the estimated convergence rate is usually too low. One example is the group of sub-Saharan African countries, meaning that the model by \citet{mankiw_1992} is not able to explain the growth in sub-Saharan Africa, because its economic fundamentals incorporated in the model are not as bad as their actual performance. The result is that, if an additional variable is added, that only indicates the membership to sub-Saharan Africa, namely the Africa Dummy, it has a significant coefficient with a negative sign. As African countries started with a lower level of income, they should converge to the income observed in regions that have similar characteristics. The presence of a negative Africa-Dummy indicates that this might not be the case. 

However, although several of the above mentioned papers seem to have succeeded in explaining the reasons for the Africa-Dummy, they did so by quite - if not completely - different arguments,
respectively its corresponding variables. In fact, adding variables to the growth regression in order to explain the Africa-Dummy is critical. In almost all these cases these additionally 
included variables just 
identified almost uniquely the belonging of a country 
to Sub-Saharan Africa, and therefore act just like the Africa-Dummy. 
For example \citet{levine_1992} test the causality of different explanatory variables in growth regressions. They summarize that most of the included variables are not robust and depend on the model. \citet{collier_1999} note that this adding of explanatory variables transfers the puzzle elsewhere. Moreover, many explanatory variables that are added in growth regressions do not necessarily identify drivers for growth. Instead, they are somehow correlated to what is not explained by the growth model but - like here - just identify a geographical region.
Finally, many country specific characteristics are time invariant so that they have already been 
accounted for in fixed effect panel models.

The naive way in which explanatory variables are added or deleted from growth models motivates to only use the explanatory variables given by \citet{mankiw_1992} and to accept that growth
for (Sub-Saharan) Africa is different. First we discuss how to identify and estimate the Africa-Dummy. When \citet{hoeffler_2002} tried to address this problem she found that the significance of the Africa-Dummy disappeared when applying a two step system GMM. 
We will briefly discuss the advantages of our approach over two-step procedures and some of the  
disadvantages of the system GMM. We call our approach the Two-Groups Least-Square Dummy-Variable estimator. This method has the advantages that it is able to estimate the Africa-Dummy in one regression step, that it is consistent even if the residuals are autocorrelated, it is able to control for all fixed effects without the need of equal variances of the fixed effects, and it gives correct standard errors and correlations for all estimated coefficients. Estimating the coefficients of the growth model with the Two-Groups Least-Square Dummy-Variable estimator identifies a negative significant Africa-Dummy. The correlations of estimates tell
us its relationship to the other returns. In fact, this punishment for Sub-Saharan African economies increases if the return to investment in physical capital decreases, if the return the depreciation rate increases or if the return to school attainment increases. We check that the Africa-Dummy is not a result of misspecification of the functional structure like nonlinearities or interactions. It does not disappear when applying a semiparamteric extension of the Two-Groups Least-Square Dummy-Variable estimator. When adding interaction effects one can observe that Sub-Saharan Africa  have had positive returns to population growth and faster convergence,
so that the Africa-Dummy becomes even significantly positive. Based on our method we can also study the evolution of the Africa-Dummy over time. Assuming world-wide similar returns as in the original model, the main finding here is that the African countries have been catching up so that this dummy has become insignificant in the recent years.

The rest of the paper is organized as follows. We first report the data selection and preparation for our study. Afterward, in Section \ref{sec3} we dedicate one section to the introduction and discussion of 
modeling and the estimation method we propose. Note that this method can equally well be applied
to identify and estimate any time invariant impact in fixed effect panel models but is not specific to the problem of studying the Africa-Dummy. All empirical findings are presented and interpreted 
in Section \ref{sec-emp}. Finally, Section \ref{sec5} concludes.

\section{Data Selection and Preparation for the Growth Model}\label{sec2}

The objective was to collect long time-series for as many countries as possible for which we can guarantee good data quality. The information sources for the empirical investigation are the Penn World Table 6.3 (PWT), World Bank's World Development indicators and \citet{barro_2010}. Except of population growth and human capital, all data come from the PWT. It collects a broad range of macroeconomic time-series for almost all countries published by \citet{heston_2009}. The beginning of a widespread availability is 1960. Most variables are published until 2007, so that observations are obtained for 48 periods. \citet{heston_2009} introduced a country rating system based on the number of participations in worldwide benchmark surveys, the variation of the accessible data and the quality of the statistical methods applied. This results in a grading scheme from A to D with descending order in which a rating of D is regarded as too weak to be included for a reliable 
empirical analysis. Furthermore, we also excluded countries that where separated in a sub-period, for example Germany and the countries of the Soviet Union. Their incorporation would have made it necessary to unify several countries to one country or to split one country in a given period in several countries. The loss of data quality when doing this is unclear. We ended up with a sample 
of complete time series of 81 countries over 48 years giving a total sample size of 3888.

The selection process might cause a sample selection bias as it results in an underrepresentation 
of African and, to some extend, Asian countries. Poor countries have weaker databases and are more likely to be excluded, but due to the inclusion of country fixed effects and the 
Africa-Dummy this can just slightly affect the slopes of the within variation. 
If this within variation is somewhat different for the under- vs the overrepresented country groups,
then there is no bias when applying our semiparametric methods which allow for flexible functional
forms. Moreover, it is clear that in our model with interactions the potential bias due to an 
underrepresentation of African countries disappears definitely by construction.  
Concerning the estimate of the Africa-Dummy it is expected that the sorting out of especially poor
countries - as it is them who have the weakest databases - will cause a positive bias, i.e.\ the punishment of being a Sub-Saharan countries might be underestimated. 
In contrast, the countries that are excluded for one of the other above mentioned reasons did 
not show any structural similarities so that it is unlikely that they cause a selection bias.
The complete list of countries included in our sample is given in Table \ref{tab:countries2}.

\begin{table}[htb]
\sffamily
\vspace{-0.25cm}
\caption{List of countries included in our sample}
\label{tab:countries2}
\begin{center}
\begin{tabular}{c|c||c|c||c|c}
Code & Country & Code & Country & Code & Country \\ \hline \hline 
ARG & Argentina & AUS & Australia &AUT  & Austria \\
BEL & Belgium & BEN & Benin & BGD & Bangladesh \\
BOL & Bolivia & BRA & Brazil & BRB & Barbados \\
BWA & Botswana & CAN & Canada & CHE & Switzerland \\
CHL & Chile & CHN & China & CMR & Cameroon \\
COG & Congo & COL & Colombia & CRI & Costa Rica \\
DNK & Denmark & DOM & Dominican Republic & ECU & Ecuador \\
EGY & Egypt & ESP & Spain & FIN & Finland \\
FJI & Fiji & FRA & France & GBR & United Kingdom \\
GHA & Ghana & GRC & Greece & GTM & Guatemala \\
HKG & Hong Kong & HND & Honduras & IDN & Indonesia \\
IND & India & IRL & Ireland & IRN & Iran \\
ISL & Iceland & ISR & Israel & ITA & Italy \\
JAM & Jamaica & JOR & Jordan & JPN & Japan \\
KEN & Kenya & KOR & Korea & LKA & Sri Lanka \\
MEX & Mexico & MLI & Mali & MUS & Mauritius \\
MWI & Malawi & MYS & Malaysia & NER & Niger \\
NGA & Nigeria & NLD & Netherlands & NOR & Norway \\
NPL & Nepal & NZL & New Zealand & PAK & Pakistan \\
PAN & Panama & PER & Peru & PHL & Philippines \\
PRT & Portugal & PRY & Paraguay & ROM & Romania \\
RWA & Rwanda & SEN & Senegal & SGP & Singapore \\
SLE & Sierra Leone & SLV & El Salvador & SWE & Sweden \\
SYR & Syria & THA & Thailand & TTO & Trinidad Tobago \\
TUN & Tunisia & TUR & Turkey & TZA & Tanzania \\
URY & Uruguay & USA & USA & VEN & Venezuela \\
ZAF & South Africa & ZMB & Zambia & ZWE & Zimbabwe
\end{tabular}
\end{center}
\vspace{-0.25cm}
\end{table}

\begin{figure}[htb]
\centering
\includegraphics[width=13.5cm]{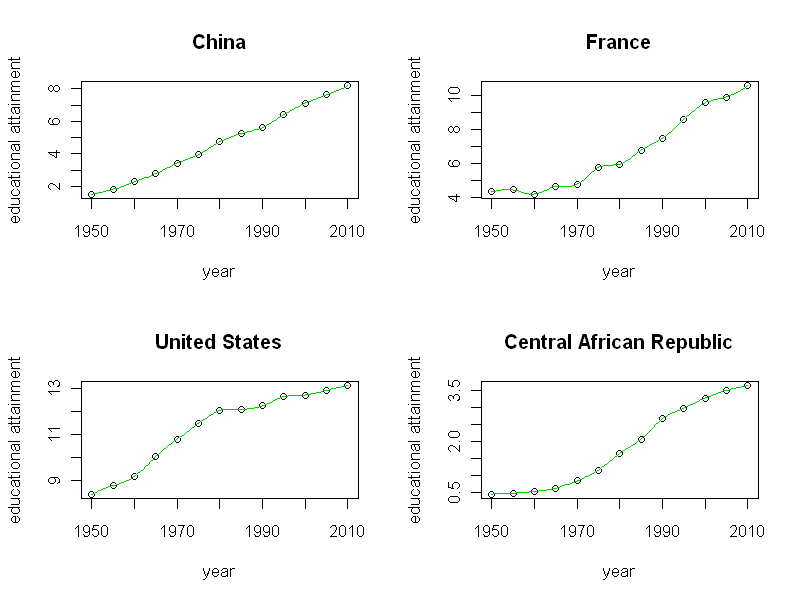}
 \vskip -0.7cm
\caption{Interpolation of schooling.}
\label{fig:ipol}
\end{figure}

Because economic growth is a consequence of changes in the production function, the output of the economy is measured as the real per worker gross domestic product (GDP). This answers the question how much each productive factor contributes on average to the growth in its country. We denote the logarithm of the per worker GDP of country $i$ at time $t$ by $y_{it}$.
The population growth refers to the working age population which is defined in the PWT as all individuals from 15 to 64 years. The data for the total population are multiplied by the share of adults in working age. We denote the growth rate of the working age population of country $i$ at time $t$ by $n_{it}$. For depreciation rates there is some accordance in the literature, see also \citet{mankiw_1992}, to expect the capital to wear out by 3\% per year and an advance in productivity of 2\% per year for all countries. We denote the logarithm of its sum plus $n_{it}$  simply by $lnn_{it}$.
The saving rate of the economy is approximated by the relative investment share of the real GDP.  
We denote the logarithm of the share of country $i$ at year $t$ by $lnsk_{it}$. 
The proxy for human capital is the educational attainment data from \citet{barro_2010} 
denoted by $lnattain_{it}$. As they were given in five years frequencies the missing values are extrapolated by interpolation splines, see the examples of China, France, US and Central African
Republic in Figure \ref{fig:ipol}. 

\begin{figure}[htb]
\centering
\includegraphics[width=13.5cm,height=10.5cm]{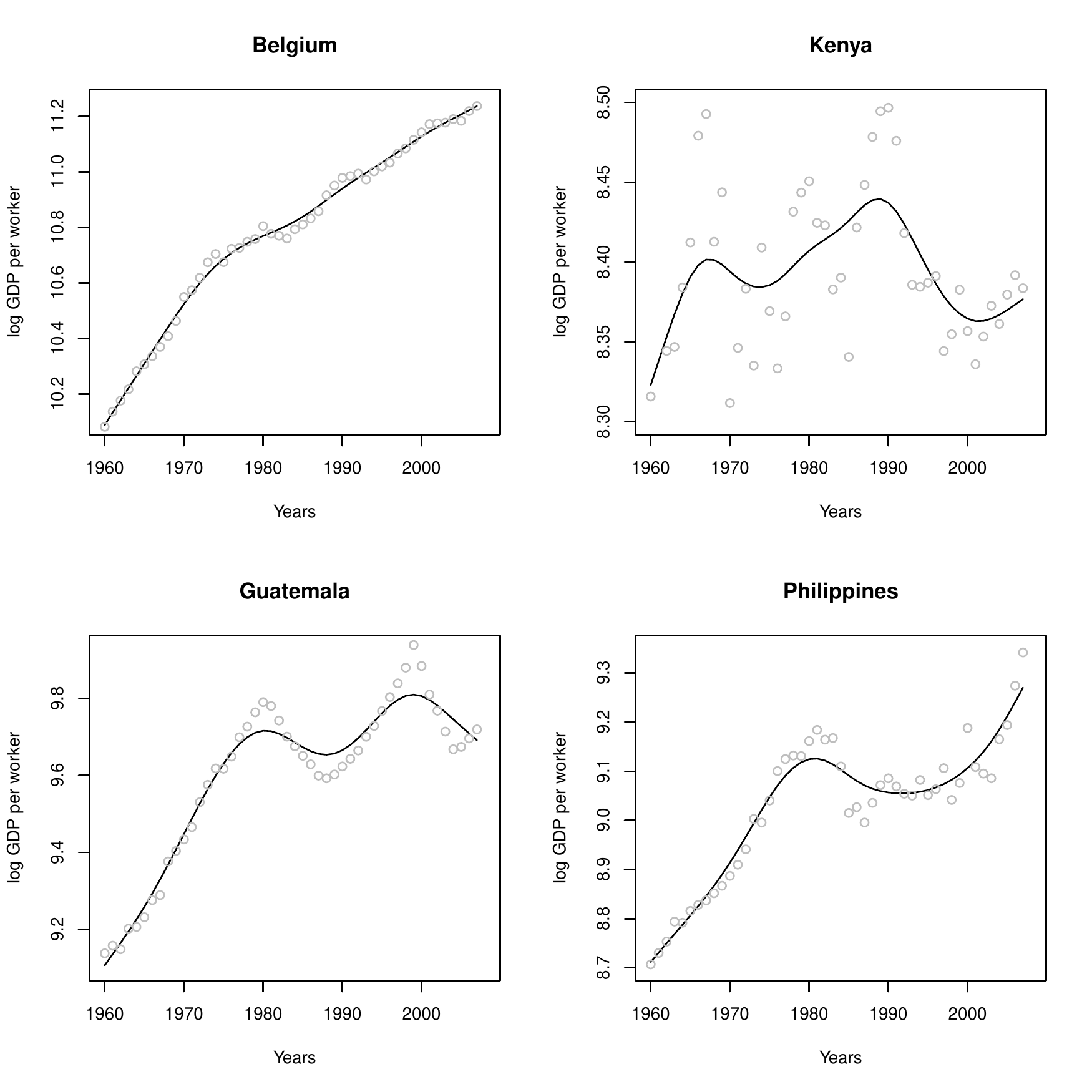}
 \vskip -0.7cm
\caption{HP Smoothing of $y_{it}$ with $\lambda=100$.}
\label{fig:yhpsmooth}
\end{figure}

\begin{figure}[htb]
\centering
\includegraphics[width=13.5cm,height=10.5cm]{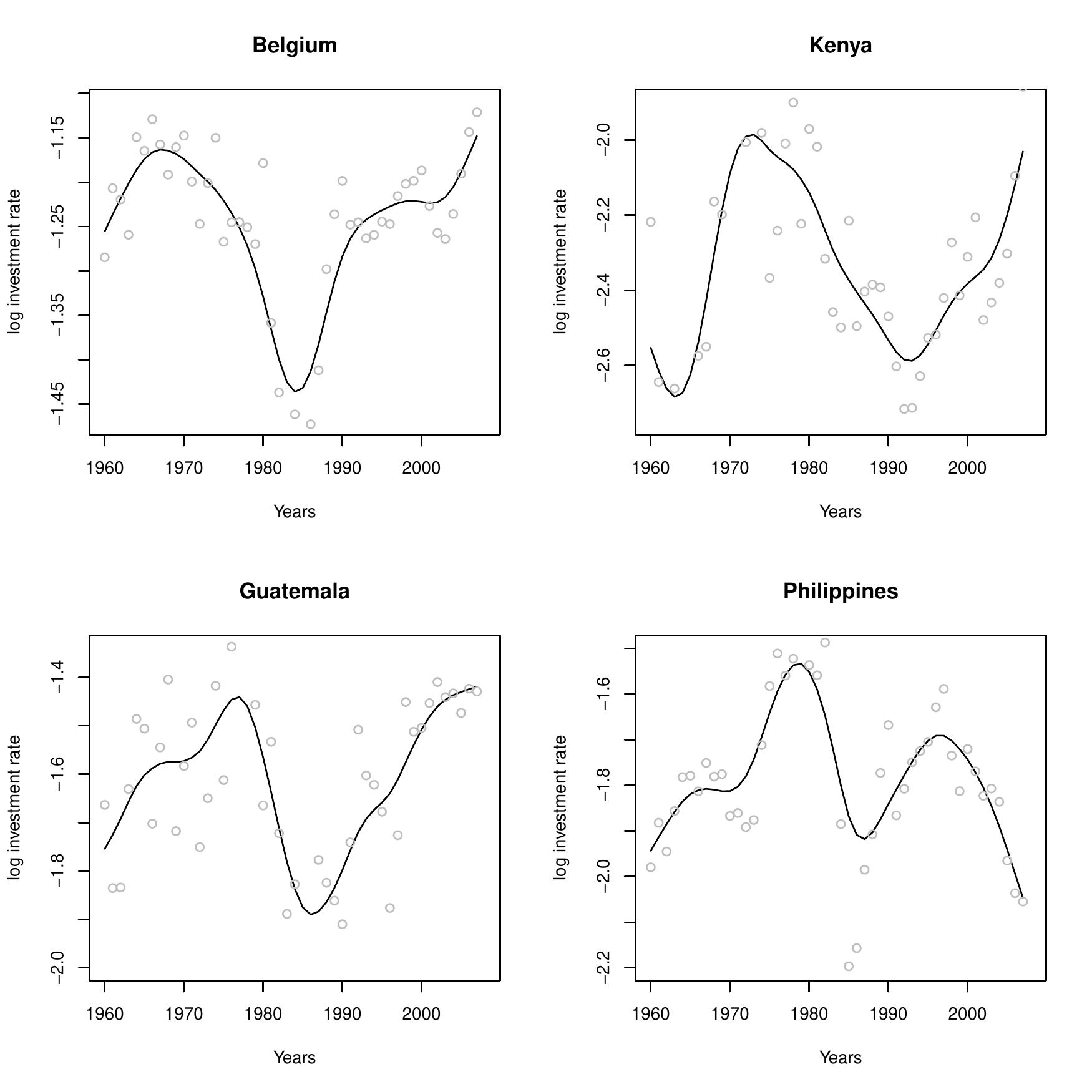}
 \vskip -0.7cm
\caption{HP Smoothing of $lnsk_{it}$ with $\lambda=25$.}
\label{fig:kihpsmooth}
\end{figure}

Most time-series have a short term cyclical component and a trend component. The Solow model addresses long run growth but not the cyclical fluctuations. Therefore, it is recommendable to 
smooth the data. As the series have different magnitudes of short term fluctuations they have to be treated in different ways. However, the series $lnn_{it}$ and $lnattain_{it}$ have only negligible short term fluctuations and are therefore not to be smoothed. The series $lnsk_{it}$ and $y_{it}$ have severe cyclical components.
We tried three possible procedures: regression smoothing, three and five years averaging, and
applying the filter of \citet{hodrick_1997}. Contrary to many other papers we finally decided for 
the last option for different reasons, see \citet{koehler_2012} for details.
The so-called HP filter decomposes a macroeconomic time-series $\tilde{\tau}_t$ in a structural trend component $\tau_t$, which accounts for sustainable long-run growth and a cyclical component $c_t$, i.e.
$$\min_{\tau_t} \sum_{t=1}^T \, (\tilde{\tau}_t-\tau_t)^2+\lambda \sum_{t=2}^{T-1}\left((\tau_{t+1}-\tau_t)(\tau_t-\tau_{t-1})\right)^2.$$
\citet{hodrick_1997} argue that $\lambda=1600$ is a reasonable choice for quarterly data which intuitively corresponds to a value of $400$ for yearly data. On the other hand, \citet{baxter_1999} argue that $\lambda$ should be chosen as the fourth power of a change in the frequency. In our case this corresponds to $6.25$. After observing the different outputs of the smoothing with different smoothing parameters, we decided to chose the compromises $\lambda=100$ for yearly growth, and $\lambda =25$ for $lnsk$, whereas the other series were already that smooth that the HP filter
for $100\ge\lambda\ge 25$ did not really change the series. 
Figure \eqref{fig:yhpsmooth} shows the smoothed series of the yearly growth rates 
and Figure \eqref{fig:kihpsmooth} for $lnsk$
 of the four countries Belgium, Kenya, Guatemala and Philippines. 

The literature deriving the augmented Solow growth model is abundant, already for the context of 
panels. Starting from the neoclassical Solow model they basically all end up with 
\begin{equation} \label{regression}
y_{it}=\rho* y_{i(t-1)} + \beta_1* lnn_{it} + \beta_2 *lnsk_{it} + \beta_3 *lnattain_{it} 
+ \eta_t + \eta_i +  \nu_{it} ,
\end{equation}
where $\eta_t$ are time fixed effects which might be skipped as argued by \citet{islam_1995}
- although he later on reincluded them to control for business cycles as he did no presmoothing -
or be modeled linearly as $\eta_t = \beta_4 t$, see \citet{sperlich_2012}. As after the HP smoothing the time trend turned out to be insignificant, we skipped it following the arguments 
of  \citet{islam_1995}.
Further,  $\eta_i$ stands e.g.\ for technical and other fixed factor differences 
like climate, land-lockedness, etc., and $\nu_{it}$ for the remaining 
unexplained heterogeneity with expectation zero. Note however, that if the $\eta_i$ are modeled 
as fixed effects without further constraints, they impede the identification of any effect coming from other time invariant factors.
Where those were of interest, people typically estimate first a fixed effects model as 
(\ref{regression}) and in a second step regress the obtained $\hat \eta_i$ on the (fixed) factors
of interest like e.g. the Africa-Dummy. Then, for further inference a correct calculation of standard errors and correlations between the estimates is necessary but typically lacking\footnote{Actually, the most common procedure in the literature is to simply believe the standard errors obtained in the second step ignoring the former step, what is just wrong.}.

\section{Identifying the Africa-Dummy} \label{sec3}

We denote the information of the dependent variable from some initial time point $1$ up to $t$ by $y_i^{t}=(y_{i1}, \ldots, y_{it})$ and the information of the exogenous variables from some initial time point $2$ up to $t$ by $x_i^t=(x_{i2}', \ldots , x_{it}')$. We assume that $\left\{\left( y_i^T, x_i^T \right) \ ,  \  i=1, \ldots , n \right\}$ are independent observations from the same probability distribution, with finite first and second order moments. 
We are aiming for estimating \eqref{regression} with the Africa-Dummy. 
If $\beta = (\beta_1 ,\beta_2 ,\beta_3 ,\beta_4)'$ and $x$ the corresponding regressors, then \eqref{regression} can be written as
\begin{equation}\label{model_gen}
y_{it}=\rho y_{i(t-1)} + x_{it}'\beta+ \eta_i + \nu_{it}.
\end{equation}
where the Africa-Dummy is a part of the country-specific effects
\begin{equation}\label{model_interest}
y_{it}=\eta_g + \rho y_{i(t-1)} + x_{it}'\beta+ SSH * 1_{SSH,i} +\tilde{\eta}_i + \nu_{it}   
\end{equation}
with $E(\tilde{\eta}_i)=0$, $1_{SSH,i}$ equals $1$ if country $i$ belongs to the group of sub-Saharan African countries and $0$ else, and $\eta_g$ is the common intercept.
We assume
\begin{equation}\label{ass_exogenity}
E(\nu_{it} |  1_{SSH,i}, y_i^{t-1}, x_i^T, \tilde{\eta}_i )=0 
\end{equation}
and 
\begin{equation}\label{ass:errors_nu}
E(\nu_{it}\nu_{js})= \left\{\begin{array}{cl} \sigma^2_{\nu}, & \mbox{if} \ i=j \ \mbox{and} \ s=t \\
0,  & \mbox{ if }  i\neq j \end{array}\right.  \ .
\end{equation}
We can relax these assumptions to allow the errors to be autocorrelated and heteroscedastic.
This might be handled then by GLS estimation or roust standard errors.
The country specific effects reflect the general productivity plus country specific characteristics 
like resources, climate, institutions, landlockedness, etc., recall discussion above.

For a vector-matrix notation we stack the time-series data, i.e.
\begin{eqnarray*}
&  y_i =(y_{i2}, \ldots , y_{iT})' \in \mathbb{R}^{T-1} \ , \quad
y_{i(-1)} =(y_{i1}, \ldots , y_{i(T-1)})' \in \mathbb{R}^{T-1}  & \\ &
 \iota =(1, \ldots , 1)' \in \mathbb{R}^{T-1}  \ , \quad
X_i =(x_{i2}, \ldots , x_{iT})  \in \mathbb{R}^{ K \times (T-1) } \ , \quad
\nu_i  =(\nu_{i2} , \ldots , \nu_{iT})' \in \mathbb{R}^{T-1} &
\end{eqnarray*}
and further
\begin{eqnarray*}
& y  =(y_1', \ldots , y_n')' \in \mathbb{R}^{n(T-1)} \ , \quad
y_{-1}  =(y_{1(-1)}', \ldots , y_{n(-1)}')' \in \mathbb{R}^{n(T-1)} & \\ & 
X =(X_1, \ldots , X_n)'  \in \mathbb{R}^{ n(T-1) \times K } \ , \quad
C =I_n \otimes \iota  \in \mathbb{R}^{n(T-1)\times n} & \\ & 
\eta =(\eta_1, \ldots , \eta_n)' \in \mathbb{R}^n \ , \quad
\nu =(\nu_1' , \ldots , \nu_n') \in \mathbb{R}^{n(T-1)} & .
\end{eqnarray*}
Equation \eqref{model_gen} can then be written as
\begin{equation} \label{equation_stacked_t} 
y= \rho y_{-1} + X \beta + C \eta + \nu \in \mathbb{R}^{n(T-1)}   
\end{equation}
and \eqref{model_interest} can be stacked in the same way. We assume without loss of generality that the data are available in the form that exactly the first $s$ rows belong to the group of sub-Saharan African countries. Denote
\begin{eqnarray*}
& 
\tilde{\eta}=(\tilde{\eta}_1, \ldots , \tilde{\eta}_n)' \in \mathbb{R}^n \ , \quad 
\iota_{n(T-1)} =(1, \ldots , 1)' \in \mathbb{R}^{n(T-1)} & \mbox{and} \\ &
\iota_{n(T-1),SSH} =(\underbrace{1, \ldots , 1}_{\in \mathbb{R}^{s(T-1)}} , \underbrace{0, \ldots , 0}_{ \in \mathbb{R}^{(n-s)(T-1)}}) \in \mathbb{R}^{n(T-1)} & .
\end{eqnarray*}
Now,  model \eqref{model_interest} can be in stacked as
\begin{equation} \label{equation_stacked_t_interest} 
y= \iota_{n(T-1)} \eta_g + \rho y_{-1} + X \beta + \iota_{n(T-1),SSH} *SSH + C \tilde{\eta} + \nu \in \mathbb{R}^{n(T-1)}. 
\end{equation}
Regression equations \eqref{model_gen} and \eqref{model_interest} have a lagged dependent variable. Therefore, including the lagged dependent variable will cause a bias when estimating the coefficients, see \citet{nickell_1981} for panels with fixed $T$. In consequence, several bias reduction procedures have been proposed, for example by \citet{kiviet_1995}, 
\citet{hahn_2002} or \citet{phillips_2007}.
We calculate the bias of Within Group estimator using the 
formulas provided in the article of \citet{phillips_2007} for different $\rho$'s.  The results show that the biases are negligible small being $<< 0.001$ for the $\beta$ estimates, and even $<< 10^{-16}$ for
the fixed effects and the Africa-Dummy.

Running the regressions using exactly \eqref{model_interest} has three drawbacks. First, the one year growth time-series shows little variation so that the coefficient of the lagged dependent variable is close to one with all other coefficients being quite small. Second, since the economy can choose its growth driving parameters simultaneously to growth, it is more natural to assume that the
drivers are the lagged values of our regressors. Third, most of the other authors considered five year time horizons taking either averaged or initial explanatory variables to represent the time horizons. So for the sake of comparison we prefer also to look at 5-year horizons. However,
taking 5-years lagged variables has two drawbacks: we move away from the situation described by \citet{mankiw_1992} and since the model deals with the evolution of the differences of the logarithms of the subsequent GDP's, 5-year horizons might generate differences that are too large to approximate growth by log-differences $\ln (GDP_t) -\ln (GDP_{t-1}) 
\approx (GDP_t-GDP_{t-1})/GDP_{t-1}$.
Therefore  we always run the regression with a one year lagged dependent variable and contemporary explanatory variables and recheck the results with a regression with a five year lagged dependent variable and five year lagged explanatory variables.

\citet{caselli_1996} applied the Difference GMM (\citet{arellano_1995}) to growth regression using linear smoothed data with five year time horizons between 1960 and 1985, but \citet{bond_2001} noted that the Difference GMM uses weak instruments because the series of the logarithms of GDP's per capita is highly persistent and recommend the System GMM. Later on, many papers have appeared using the System GMM. 
\citet{hoeffler_2002} addresses the problem of estimating the Africa-Dummy in growth regressions 
(using a two-step procedure as indicated above) coming to the conclusion that System GMM is the preferred method. As most authors use linear smoothing instead of applying the HP filter, their time-series are short which leads to few instruments. The number of instruments when having time-series data with $T=48$ is however very large causing various problems.
One general problem of GMM is a bias that occurs when too many instruments are used, see 
for example \citet{tauchen_1986} or \citet{ziliak_1997}.
Serious problems occur also when estimating the optimal weighting matrix of GMMs. 
The number of elements to be estimated is quadratic in the number of instruments and therefore quartic in $T$. Moreover, the elements of the optimal matrix are fourth moments of the underlying distributions because they are second moments of the result of differenced variables times variables. \citet{roodman_2009} notes that a common symptom for estimations of the weighting matrix is that they are singular. Therefore, the generalized inverse rather than the inverse is calculated. This can give results that are far away from the theoretical one on which further inference is built up. The breakdown tends to occur as the number of instruments approaches (from below) $n$. Note that we have 4554 instruments with only $n=81$ countries when using the System GMM. 

Different procedures were proposed to reduce the number of instruments. The Hansen J-Test (see \citet{hansen_1982}) usually checks the validity of instruments, but as for example \citet{bowsher_2002} observed, a too large number of instruments weakens the test dramatically, see also \citet{roodman_2009}. There does not exist a reliable test available that tells us how many and which instruments to choose. Finally, note that the System GMM shows serious biases when the variation of the fixed effects is larger than the residual's variance (what for macro models is almost always the case), see e.g. \citet{hayakawa_2009}, and that the required crucial initial conditions are least likely to be fulfilled in case of highly persistent time-series as in our case, see again \citet{roodman_2009}. 
All together, with the inefficiency and problems of correct inference 
of a two-step method to calculate the Africa-Dummy estimate, it is not surprising that using System GMM one obtains insignificant results. 

% discuss the idea of Hausman-Taylor for dynamic panel data models

To be able to identify and estimate \eqref{model_interest} directly, we assume that the errors of the sub-Saharan African countries sum up to zero and that the errors of the non-sub-Saharan African countries sum up to zero separately 
\begin{equation} \label{country_errors}
\sum_{i=1}^s \tilde{\eta}_i=0 \ \mbox{and} \ \sum_{i=s+1}^n \tilde{\eta}_i=0.
\end{equation}
With this assumption specify
\begin{eqnarray} \label{eq_two_gr_lsdv}
& y= \rho y_{-1} + X \beta + C_{SSH} \eta_{SSH} + \nu \in \mathbb{R}^{n(T-1)} & , 
\\
& \eta_{SSH}=(\eta_g , SSH, \tilde{\eta_1}, \ldots , \tilde{\eta}_{s-1}, \tilde{\eta}_{s+1}, \ldots , \tilde{\eta}_{n-1} )' \in \mathbb{R}^n  & \nonumber
\\[2mm] &
C_{SSH}=
 \left(
\begin{array}{*{10}{c}}  
\iota &\iota &\vline&\iota &      &      &\vline&      &				 &        \\
\vdots&\vdots&\vline&      &\ddots&			 &\vline&	    &        &				 \\
\iota &\iota &\vline&      &			 &\iota&\vline&      &        &				 \\
\iota & \iota&\vline&-\iota&\cdots&-\iota&\vline&      &        &        \\ \hline
\iota	&      &\vline&      &			&			 &\vline&\iota &	       &				 \\
\vdots&      &\vline&      &			&      &\vline&	    & \ddots &				 \\
\iota &      &\vline&      &			&      &\vline&      &        & \iota  \\
\iota &      &\vline&      &     &       &\vline&-\iota&\cdots  &-\iota
\end{array}
\right) \in \mathbb{R}^{n(T-1) \times n},
&  \nonumber
\end{eqnarray}
where the lower right box refers to the non-sub-Saharan African countries and has $n-s-1$ columns and $(n-s)(T-1)$ rows and the upper middle box refers to the sub-Saharan African countries and has $s-1$ columns and $s(T-1)$ rows. It is easy to check that
\begin{eqnarray*}
& C_{SSH}'C_{SSH}=(T-1)
 \left(
\begin{array}{*{3}{c}}  
Z_1 &   &    \\
    &Z_2&    \\
    &   & Z_3
\end{array}
\right) \in \mathbb{R}^{n \times n}, \quad 
Z_1=
 \left(
\begin{array}{*{2}{c}}
n & s \\
s & s 
\end{array}
\right)  ,  
& \\[2mm] &
Z_2=
 \left(
\begin{array}{*{4}{c}}
2       & 1        & \ldots  & 1         \\
1       & 2        & \ddots  & \vdots     \\ 
\vdots  & \ddots   & \ddots  & 1       \\
1       & \ldots   &  1      &  2      
\end{array}
\right) \in \mathbb{R}^{(s-1) \times (s-1)},
\quad
Z_3=
\left(
\begin{array}{*{4}{c}}
2       & 1        & \ldots  & 1         \\
1       & 2        & \ddots  & \vdots     \\ 
\vdots  & \ddots   & \ddots  & 1       \\
1       & \ldots   &  1      &  2      
\end{array}
\right) \in \mathbb{R}^{(n-s-1) \times (n-s-1)} .
\end{eqnarray*}
The inverses of $Z_1$, $Z_2$ and $Z_3$ exist and are given by
$$
Z_1^{-1}=\frac{1}{n-s}
 \left(
\begin{array}{*{2}{c}}
1 & -1 \\
-1 & n/s 
\end{array}
\right) \in \mathbb{R}^{2 \times 2},
$$
$$
Z_2^{-1}=\frac{1}{s}
 \left(
\begin{array}{*{4}{c}}
(s-1)     & -1         & \ldots  &    -1       \\
-1        & (s-1)      & \ddots  &    \vdots   \\ 
\vdots    &  \ddots    & \ddots  &    -1       \\
-1        &    \ldots  &  -1     &  (s-1)   
\end{array}
\right) \in \mathbb{R}^{(s-1) \times (s-1)},
$$ 
and 
$$
Z_3^{-1}=\frac{1}{n-s}
 \left(
\begin{array}{*{4}{c}}
(n-s-1)   & -1         & \ldots  &    -1       \\
-1        & (n-s-1)    & \ddots  &    \vdots   \\ 
\vdots    &  \ddots    & \ddots  &    -1       \\
-1        &    \ldots  &  -1     &  (n-s-1)   
\end{array}
\right) \in \mathbb{R}^{(n-s-1) \times (n-s-1)}.
$$ 
It follows that  
$$(C_{SSH}'C_{SSH})^{-1}=\frac{1}{T-1}
 \left(
\begin{array}{*{3}{c}}  
Z_1^{-1}  &        &         \\
          &Z_2^{-1}&         \\
          &        & Z_3^{-1}
\end{array}
\right) \in \mathbb{R}^{n \times n}.
$$ 
Note that the existence of $(C_{SSH}'C_{SSH})^{-1}$ is equivalent to that the columns of $C_{SSH}$ are linear independent, meaning that the model can be identified. It is now easy to check that 
$$M_{C_{SSH}}=I_{n(T-1)}-C_{SSH}(C_{SSH}'C_{SSH})^{-1} C_{SSH}'=I_{n(T-1)}-I_n \otimes \iota \iota' \in \mathbb{R}^{n(T-1) \times n(T-1)}.$$
Therefore, $\rho$ and $\beta$ can be estimated by the Within Group estimator. Furthermore,
$$\hat{\eta}_{SSH}=(C_{SSH}'C_{SSH})^{-1}C_{SSH}'(y-\hat{\rho}_{WG} y_{-1} - X \hat{\beta}_{WG}).$$
Solving this gives the Two-Groups Least-Square Dummy-Variable estimator
\begin{equation}\label{lsdv_ssh}
\begin{split}
&\hat{\rho}=\hat{\rho}_{WG}, \ \hat{\beta}=\hat{\beta}_{WG}, \ \hat{\eta}_g=\bar{\eta}_{NA}, \ \hat{SSH}=\bar{\eta}_{A}- \bar{\eta}_{NA},\\
&\hat{\tilde{\eta}}_j=\bar{\eta}_{j}-\bar{\eta}_{A} \ \mbox{for } j \in \left\{1, \ldots , s-1 \right\} \mbox{and} \ \hat{\tilde{\eta}}_j=\bar{\eta}_{j}-\bar{\eta}_{NA} \ \mbox{for } j \in \left\{s+1, \ldots , n-1 \right\}.
\end{split}
\end{equation}
With \eqref{lsdv_ssh} and $-\tilde{\eta}_1- \ldots - \tilde{\eta}_{s-1}=\tilde{\eta}_s$ we have $\hat{\tilde{\eta}}_s= \bar{\eta}_s-\bar{\eta}_{A}$ and in the same manner $\hat{\tilde{\eta}}_n=\bar{\eta}_n-\bar{\eta}_{NA}$.
The total country-specific effect of a sub-Saharan African country with index $j \in \left\{1, \ldots ,s\right\}$ is $\hat{\eta}_g+\hat{SSH} + \hat{\tilde{\eta}}_j =\bar{\eta}_j$ and that of a non-sub-Saharan African country with index $j \in \left\{s+1, \ldots ,n\right\}$ is $\hat{\eta}_g+ \hat{\tilde{\eta}}_j=\bar{\eta}_j$. 
The Two-Groups Least-Square Dummy-Variable estimator allows to reliably estimate the correlations of the Africa-Dummy to other regressors. Furthermore, as it does not use the inefficient Instrumental Variable method, it is more efficient. Another example of the Least-Squares method is that it remains being consistent even if the residuals are heteroscedastic and serially correlated.

\section{Empirical Results for different Specifications} \label{sec-emp}

Table \eqref{tab:coefficients_2} show the estimated coefficients with its standard errors. The
outcome and interpretation of the five year lagged model is similar to that of the one year lagged model for all estimation methods. The coefficient of $lnn$ is almost zero in the one year lagged model and at least becomes negative significant on ten percent level in the five year lagged model. 
We will see later that this is due to the heterogeneity between the different regions.
As often observed, the coefficient of $lnattain$ is negative. Note that the indicator by \citet{barro_2010} does not take the quality of schooling into account. Further, even when 
their data were corrected for the account change after 2005 (in some countries) the school attainment increases for almost all countries while the growth rate does not. The numerical consequence is a negative coefficient\footnote{Recall that we are estimating only the impact of 
country specific (i.e. within) variation.}.

\begin{table}[htb]
\sffamily
\vspace{-0.25cm}
\begin{center}
\caption{Two Groups Fixed Effects Estimates with standard errors}
\label{tab:coefficients_2}
 \begin{tabular}{*{5}{l}}
 \hline
  & \multicolumn{2}{c}{one year lag} & \multicolumn{2}{c}{five years lag}  \\  \hline\hline
  Intercept   &   0.1795*** &   (0.0117)   & 1.1343***   & (0.0635) \\
  lag y       &    0.9897***  &    (0.0011)  & 0.8926***  &(0.0061) \\
  lnn         &    0.0008 &    (0.0025)      & -0.0240  (0.0127) \\
  lnsk        &   0.0275*** &    (0.0012)   & 0.0813***  & (0.0063) \\
  lnattain    &  -0.0150*** &   (0.0010)  & -0.0493***   &  (0.0053) \\
  SSH         & -0.0109***  &   (0.0017)   & -0.1551***   & (0.0090) \\  \hline\hline
* $p:\; \le 0.05$ &**$\le 0.01$&***$\le 0.001$
\end{tabular}
\end{center}
\vspace{-0.25cm}
\end{table}

When next looking at the correlation of the coefficients we see that 
the Africa-Dummy is larger, the smaller the coefficient of $lnn$ or $lnattain$ are or the larger the coefficient of $lnsk$ is. Nevertheless, its correlations to the coefficient of $lnattain$ and $lnn$ are small. In other words, if the return to investment in physical capital increases, the punishment of belonging to sub-Saharan Africa decreases. Furthermore, if the return to the depreciation rate or the school attainment increases, the punishment of belonging to Sub-Saharan Africa increases. 

\begin{table}[htb]
\sffamily
\vspace{-0.25cm}
\begin{center}
\caption{Correlation of Africa-Dummy with other coefficient estimates}
\label{tab:correlations}
 \begin{tabular}{*{6}{l}}
 \hline
  Model    &Corr lnn & Corr lnsk & Corr lnattain  \\         
\hline \hline
 one year lag   & -0.1170  & 0.5641   &  -0.0938  \\
 five years lag & -0.1279  & 0.5252   &  -0.0537  \\
\hline
\end{tabular}
\end{center}
\vspace{-0.25cm}
\end{table}

\begin{figure}[htb]
\centering
\includegraphics[width=12cm]{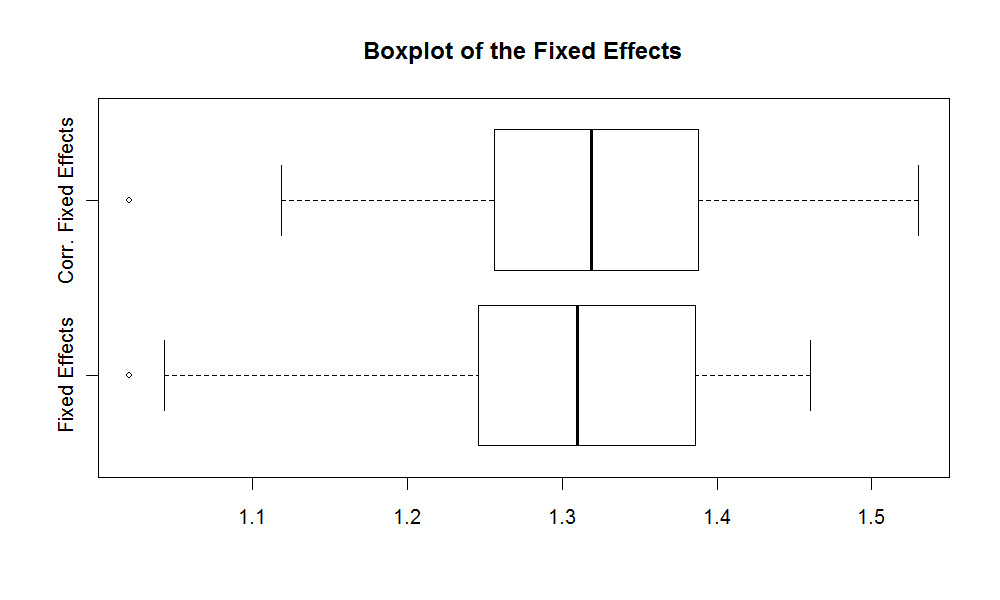}
 \vskip -0.70cm
\caption{Boxplot of the fixed effects for the five year lagged model.}
\label{fig:boxplot_country_err_5}
\end{figure}

The Two-Groups Least-Square Dummy-Variable estimator is able to estimate the decomposition $\tilde{\eta}_i+\eta_g + SSH* 1_{SSH;i}$ .  We denote $\tilde{\eta}_i+\eta_g + SSH* 1_{SSH;i}$  by fixed effects and $\tilde{\eta}_i+\eta_g $ by corrected fixed effects. The corrected fixed effects are larger than the fixed effects in case of a sub-Saharan African country and equal for all other countries. Figure \eqref{fig:boxplot_country_err_5} % \eqref{fig:boxplot_country_err_1} 
shows boxplots of the fixed effects in the five years lagged model. We observe that both distributions are slightly skewed to the left.  The tails of the corrected fixed effect support a symmetric distribution but as the median is closer to the first quartile than to the third quartile, the distribution is also slightly skewed to the left. 

Let us turn to the semiparametric modeling to account for possible functional misspecification.
The growth model by \citet{mankiw_1992} suggests the regression equation \eqref{model_interest} which has a linear functional structure. We investigate if a misspecification of this functional structure is responsible for that the Africa-Dummy is negative and significant. Note that 
functional misspecification causes biased coefficient estimates similar to those of
the biases of omitted variable, see \citet{koehler_2012} for more details.
We use cubic B-Splines of degree three with equidistant knots to relax the functional structure of the variables $lnn$, $lnsk$ and $lnattain$. The number of knots has been chosen in a way that takes the sample size as well as the number of regressors into account. Akaike's Information Criterion results in choosing models with too many parameters when having large samples. The Bayesian Information Criterion punishes harder for choosing a lot of explanatory variables. Therefore, we chose the number of knots with respect to that it minimizes the Bayesian Information Criterion. More precisely, we vary the number of knots between three and ten and choose the combination that minimizes the Bayesian Information Criterion. The result for the one year lagged model is zero knots for the variables $lnn$, $lnattain$, and one knot for the variable $lnsk$. The result for the five year lagged model is one knot for all variables. When running these regressions we observe that the coefficient of the lagged dependent variable increases from $0.9897$ to $ 0.9920$ in the one year lagged model and decreases from $0.8926$ to $ 0.8911$ in the five year lagged model. The intercept decreases from $0.1905$ to $0.0322$ in the one year lagged model and from $1.2894$ to $ 0.8834$ in the five year lagged model. The magnitude of the Africa-Dummy increases slightly from $-0.0109$ to $-0.0113$ in the one year lagged model and from $-0.1551$ to $-0.1582$ in the five year lagged model. In both cases we observe a highly significant Africa-Dummy, i.e.\ the significance of the Africa-Dummy cannot be explained by a misspecification of the functional structure
in an additive model.

We next turn to the question of potential interaction effects and consider model \eqref{model_interest} with interactions effects that allow for time varying punishments of sub-Saharan African countries, i.e.\ we add $x_{it} 1_{SSH,i}$ to equation \eqref{model_interest}.
The results are given in Table \eqref{tab:coefficients_interaction}.
There we observe a positive significant interaction effect of the coefficient of $lnn$, where
now, cf.\ Table \eqref{tab:coefficients_interaction}, the return to $lnn$ is significant
negative as predicted by economic growth theory.  
For the one year lagged model the total coefficient of $lnn$ is $-0.0129+0.0357=0.0228$ and for the five year lagged model $-0.0760+0.1535=0.0775$, i.e.\ in Africa population growth has a positive
impact on growth what is probably not that surprising given the dominance of the manufacturing 
and (mainly man power based) agricultural sectors. The negative interaction with the base
GDP (in the five years lag model) indicates that African countries converge faster when
controlling for the interaction with $lnn$. Further interactions are insignificant except with 
$lnattain$ in the one-year lag model at the 5\% level. 
Surprisingly, having controlled for the specific African feature of positive impact of population growth on GDP, the Africa-Dummy is significantly  positive. All in all, accounting for the special (higher) returns to population growth in Africa we find faster beta-convergence and higher 
conditional growth for Sub-Saharan African countries.

\begin{table}[htb]
\sffamily
\vspace{-0.25cm}
\begin{center}
\caption{Estimates and standard errors of the growth regression with interactions}
\label{tab:coefficients_interaction}
 \begin{tabular}{*{5}{l}}
 \hline & \multicolumn{2}{c}{one year lag}  & \multicolumn{2}{c}{five years lag} \\ \hline\hline
  Intercept   &    0.1588*** &    (0.0134)   &    1.0938***  &   (0.0724)    \\
  SSH         &     0.0646*  &    (0.0266)  &    0.6151***   &   (0.1451)     \\
  lag y       &    0.9895*** &   (0.0013)   &    0.8976***    &   (0.0070)    \\
 $1_{SSH}$ lag y  &   0.0020    &   (0.0027)    &    -0.0397**   &    (0.0147)    \\            
  lnn         &   -0.0129*** &   (0.0031)    &    -0.0760***   &    (0.0159)     \\
 $1_{SSH}$ lnn    &  0.0357***   &   (0.0052)   &   0.1535***    &   (0.0265)     \\
  lnsk        &    0.0268*** &   (0.0016)     &    0.0752***   &   (0.0081)    \\
 $1_{SSH}$ lnsk   &   0.0028    &   (0.0025)     &    0.0145     &   (0.0129)      \\
  lnattain    &    -0.0175***  &   (0.0013)  &  -0.0498***   &   (0.0070)   \\
 $1_{SSH}$ lnattain &  0.0047*  &   (0.0020)     &   0.0017  &  (0.0108)    \\  \hline\hline
& \multicolumn{4}{c}{*$p:\; \le 0.05$\ \ **$\le 0.01$\ \ ***$\le 0.001$}
\end{tabular}
\end{center}
\vspace{-0.25cm}
\end{table}

% \section{The Development of the Africa-Dummy over Time}\label{time_development}

To investigate how the Africa-Dummy evolves over time consider model
\begin{equation}\label{model_time}
y_{it}= \eta_g + \rho y_{i(t-1)} + x_{it}'\beta + \sum_{s=2}^T SSH_s*d_{SSH,t}(i,s) + \tilde{\eta}_i + \nu_{it},
\end{equation}
with $t=2, \ldots , T$ and $i=1, \ldots , n$, where $d_{SSH,t}(i,s)=1$ if country $i$ belongs to sub-Saharan Africa and $s=t$ and $d_{SSH,t}(i,s)=0$ else, and still $\sum_{i=1}^s \tilde{\eta}_i=0$,
  $\sum_{i=s+1}^n \tilde{\eta}_i=0$ for identifying the model. Stacking first time-series and then cross-sectional data yields
$$y=\rho y_{-1} + X\beta + (\iota_{SSH} \otimes I_{T-1}) SSH +C \eta + \nu \in \mathbb{R}^{n(T-1)},$$
where $SSH=(SSH_2, \ldots , SSH_T)' \in \mathbb{R}^{T-1}$ , $\eta=(\eta_g, \tilde{\eta_1}, \ldots , \tilde{\eta}_{s-1}, \tilde{\eta}_{s+1}, \ldots , \tilde{\eta}_{n-1} )' \in \mathbb{R}^{n-1},$   
$$
C=
 \left(
\begin{array}{*{9}{c}}  
\iota &\vline&\iota &      &      &\vline&      &				 &        \\
\vdots&\vline&      &\ddots&			 &\vline&	    &        &				 \\
\iota &\vline&      &			 &\iota&\vline&      &        &				 \\
\iota &\vline&-\iota&\cdots&-\iota&\vline&      &        &        \\ \hline
\iota	&\vline&      &			&			 &\vline&\iota &	       &				 \\
\vdots&\vline&      &			&      &\vline&	    & \ddots &				 \\
\iota &\vline&      &			&      &\vline&      &        & \iota  \\
\iota &\vline&      &     &       &\vline&-\iota&\cdots  &-\iota
\end{array}
\right) \in \mathbb{R}^{n(T-1) \times (n-1)}.
$$
Note that this matrix does not contain the time varying Africa-Dummies. The lower right box refers to the non sub-Saharan African countries and has $n-s-1$ columns and $(n-s)(T-1)$ rows, the upper middle box refers to the sub-Saharan African countries and has $s-1$ columns and $s(T-1)$ rows and the first column refers to the intercept. The complete dummy matrix with the Africa-Dummies is $(\iota_{SSH} \otimes I_{T-1}, C) \in \mathbb{R}^{n(T-1) \times (n+(T-1))}$ and has full column rank. In the same way we formulate the five years lag model 
$$y_{it}=\eta_g + \rho y_{i(t-5)} + x_{i(t-5)}'\beta + \sum_{s=6}^T SSH_s*d_{SSH,t}(i,s) + \tilde{\eta}_i + \nu_{it}  .$$

\begin{table}[htb]
\sffamily
\vspace{-0.25cm}
\begin{center}
\caption{Coefficient estimates and standard errors for a model with time-varying Africa-Dummy}
\label{tab:coefficients_time}
 \begin{tabular}{*{5}{l}}
 \hline
      & \multicolumn{2}{c}{one year lag}  & \multicolumn{2}{c}{five years lag}  \\ \hline\hline
  Intercept   &    0.1832*** &    (0.0117)  &   1.2654***    &   (0.0636)   \\
  lag y       &    0.9911***  &  (0.0011)   &    0.8964***    &   (0.0062)    \\
  lnn         &   0.0012   &   (0.0025)     &    -0.0214     &    (0.0128) \\
  lnsk        &    0.0277***  &   (0.0013)   &    0.0834***     &   (0.0065)   \\
  lnattain    &    -0.0175***  &   (0.0012)   &    -0.0510***  &   (0.0065)  \\ \hline\hline
&\multicolumn{4}{c}{* $p:\; \le 0.05$ \ \ **$\le 0.01$ \ \ ***$\le 0.001$}
\end{tabular}
\end{center}
\vspace{-0.25cm}
\end{table}

\begin{figure}[htb]
\centering
\includegraphics[width=12cm]{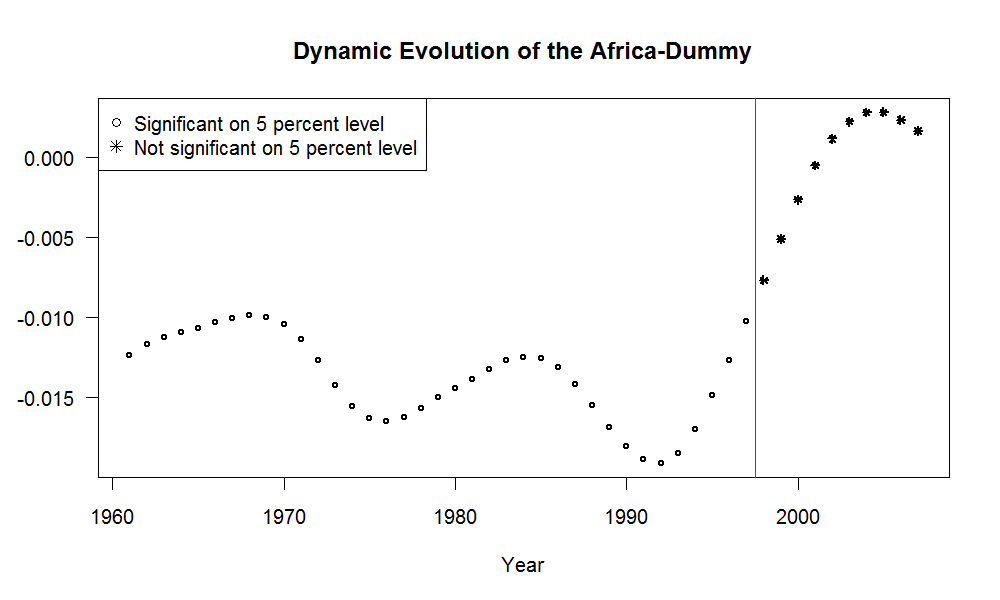}
\vskip -0.7cm
\caption{The Evolution of the Africa-Dummy in the one year lagged model}
\label{fig:ad_time_1}
\end{figure}

\begin{figure}[htb]
\centering
\includegraphics[width=12cm]{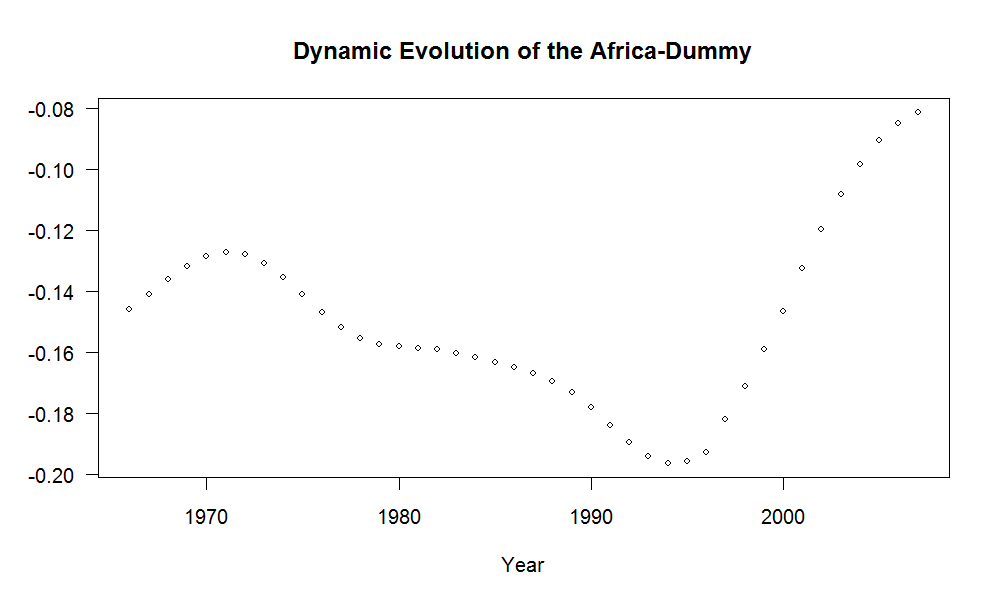}
\vskip -0.7cm
\caption{The Evolution of the Africa-Dummy in the five year lagged model}
\label{fig:ad_time_5}
\end{figure}

The results for the estimators of the coefficients are given in Table \eqref{tab:coefficients_time}. We observe that the estimators of the coefficients of \eqref{model_time} are similar to the model with a static Africa-Dummy, equation \eqref{model_interest}. Figures \eqref{fig:ad_time_1} and \eqref{fig:ad_time_5} show that the Africa-Dummy varies a lot over time. Apart from the
downward bumps during the oil crises in the mid 70s and the end of the cold war [together with
the break down of the Eastern block economies] it had a general though small downward trend
but then started to strongly increase since the mid-nineties. When considering the one year lagged model it even becomes insignificant
about the turn of the Millennium. However, looking at the five years growth the Africa-Dummy is still significant though the punishment has decreased from about $-0.20$ to $-0.08$. This is in 
accordance with our findings from the exercise with interaction terms. There we found a
positive impact on conditional growth and that Sub-Saharan African countries have been converging somewhat faster in the last 50 years compared to the rest of the world.  
% It would be interesting to find out whether also the returns to $lnn$ have been converging world wide.

\section{Conclusion}\label{sec5}

By smoothing the data with the Hodrick-Prescott filter we obtain yearly time-series that represent the connection of one time-series of an economy to another. When doing this, the length of the time-series is sufficiently large, so that the Nickel bias that may appear in a dynamic panel  growth regression is negligibly small. The centering of the country specific fixed effects allows to identify and estimate the Africa Dummy directly in the classical growth model. 
This entails several advantages over the
else so far used methods. Then, estimating the coefficients of the growth regression with the Two-Groups Least-Square Dummy-Variable estimator identifies a negative significant Africa-Dummy. The analysis of correlations of coefficient estimates reveals that this handicap for Sub-Saharan African economies increases if the return to investment in physical capital decreases, if the return the depreciation rate increases, or if the return to school attainment increases.

The Two-Groups Least-Square Dummy-Variable estimator is also used to relax the functional structure of the growth regression equation. We observe that the significance of the Africa-Dummy does not disappear when applying a semiparametric model so that it cannot be explained by a misspecification of the functional form. In contrast, when modeling the returns more flexibly we observe that Sub-Saharan African countries have clearly positive returns to the population growth rate,
exhibit faster beta-convergence, such that the pure Africa-Dummy becomes even positive. 
As discussed, all these findings make sense but were hidden in former studies due to improper
modeling and estimation methods. The imposing of equal returns for all regions in a world sample 
forces the Africa-Dummy to correct the mean, resulting in a negative coefficient. 

Having seen that allowing for heterogeneous returns exhibits faster conditional growth
and convergence for Sub-Saharan Africa, it is not surprising that 
when we estimate the evolution of the Africa-Dummy by our extended version of the Two-Groups Least-Square Dummy-Variable estimator, one can see how this gap diminishes. It can clearly be 
observed how the Africa-Dummy changes over time being strongly 
increasing since the mid-nineties. In the one year lag model it has even become insignificant 
in the recent years.

\end{document}